\documentclass[aps,prl,amsmath,twocolumn,superscriptaddress,showpacs]{revtex4-1}

\usepackage{graphicx}
\usepackage{color}
\newcommand{\Fref}[1]{Figure~\ref{#1}}
\newcommand{\fref}[1]{Fig.~\ref{#1}}

\newcommand{\bak}{(Ba$_{1-x}$K$_x$)Fe$_2$As$_2$}
\newcommand{\spm}{$s_{\pm}$}

\begin{document}

\title{Energy gap evolution across the superconductivity dome \\ in single crystals of (Ba$_{1-x}$K$_x$)Fe$_2$As$_2$}

\author{Kyuil Cho}
\affiliation{Ames Laboratory, Ames, IA 50011, USA}
\affiliation{Department of Physics and Astronomy, Iowa State University, Ames, IA 50011, USA}

\author{Marcin Ko\'nczykowski}
\affiliation{Laboratoire des Solides Irradiés, Ecole Polytechnique, CNRS, CEA, Université Paris-Saclay, 91128 Palaiseau Cedex, France}

\author{Serafim Teknowijoyo}
\affiliation{Ames Laboratory, Ames, IA 50011, USA}
\affiliation{Department of Physics and Astronomy, Iowa State University, Ames, IA 50011, USA}

\author{Makariy A. Tanatar}
\affiliation{Ames Laboratory, Ames, IA 50011, USA}
\affiliation{Department of Physics and Astronomy, Iowa State University, Ames, IA 50011, USA}

\author{Yong Liu}
\affiliation{Ames Laboratory, Ames, IA 50011, USA}

\author{Thomas A. Lograsso}
\affiliation{Ames Laboratory, Ames, IA 50011, USA}
\affiliation{Department of Materials Science and Engineering, Iowa State University, Ames, IA 50011, USA}

\author{Warren E. Straszheim}
\affiliation{Ames Laboratory, Ames, IA 50011, USA}

\author{Vivek Mishra}
\affiliation{Joint Institute for Computational Sciences, University of Tennessee, Knoxville, TN 37996, USA}
\affiliation{Center for Nanophase Materials Sciences, Oak Ridge National Laboratory, Oak Ridge, TN 37831, USA}

\author{Saurabh Maiti}
\affiliation{Department of Physics, University of Florida, Gainesville, Florida 32611, USA}

\author{Peter J. Hirschfeld}
\affiliation{Department of Physics, University of Florida, Gainesville, Florida 32611, USA}

\author{Ruslan Prozorov}
\email[Corresponding author: ]{prozorov@ameslab.gov}
\affiliation{Ames Laboratory, Ames, IA 50011, USA}
\affiliation{Department of Physics and Astronomy, Iowa State University, Ames, IA 50011, USA}


\begin{abstract}
The mechanism of unconventional superconductivity in iron-based  superconductors (IBSs) is one of the most intriguing questions in current materials research. Among non-oxide IBSs, (Ba$_{1-x}$K$_x$)Fe$_2$As$_2$ has been intensively studied because of its high superconducting transition temperature and  fascinating evolution of the superconducting gap structure from being fully isotropic at optimal doping (\(x\approx\)0.4) to becoming nodal at $x > $0.8. Although this marked evolution was identified in several independent experiments, there are no details of  the gap evolution to date because of the lack of high-quality single crystals covering the entire K-doping range of the superconducting dome. We conducted a systematic study of the London penetration depth, $\lambda (T)$, across the full phase diagram for different concentrations of point-like defects introduced by 2.5 MeV electron irradiation. Fitting the low-temperature variation with the power law, $\Delta \lambda \sim T^{n}$, we find that the exponent $n$ is the highest and $T_c$ suppression rate with disorder  is the smallest at optimal doping, and they evolve with doping being away from optimal, which is consistent with increasing gap anisotropy, including an abrupt change around $x\simeq 0.8$, indicating the onset of nodal behavior. Our analysis using a self-consistent $t$-matrix approach suggests the ubiquitous and robust nature of  s$_{\pm}$  pairing in IBSs and argues against a previously suggested transition to a $d-$wave state near $x=1$ in this system.\\

Published: ScienceAdvances \textbf{2} (9), e1600807, 30 September 2016.

DOI: 10.1126/sciadv.1600807
\end{abstract}

\maketitle

\section{Introduction}
Understanding the mechanisms of superconductivity in iron-based superconductors (IBSs) is a challenging task, partially  due to the multiband character of interactions and scattering \cite{Chubukov2012ARCMP,Chubukov-Hirschfeld-PT_2015,BaRb122uSR,Hirschfeld2016} .
On the other hand, the rich chemistry of IBSs offers a unique opportunity to study the physics within one family of materials and test material-specific theories of superconductivity. It is widely believed that spin fluctuations due to repulsive Coulomb interactions are responsible for superconductivity and lead to sign-changing pair states around the Fermi surface.  Theories of superconductivity based on exchange of these electronic excitations predict that large-momentum pair scattering processes dominate the pairing interactions, but details distinguish between competing pair states, usually $s-$wave and $d-$wave.  For the simplest band structures characteristic of these systems, it was found that optimally doped systems should have a fully gapped, $s-$wave ground state, but as the system was overdoped $d-$wave would become more competitive and even the $s-$wave state would become extremely anisotropic \cite{Maiti2011PRB}.  Systematically testing these predictions would be an important step toward understanding the origins of superconductivity in these systems.

Among various IBSs, \bak~ (BaK122) is, perhaps, one of the most interesting and intensively studied compounds, exhibiting an unusual
variation of the superconducting gap structure across the superconducting dome that exists between \(x\approx0.18\) and \(1\). In the optimally-doped region, $x \approx 0.35$ to $0.4$, two effective  isotropic superconducting gap scales (roughly with a 2:1 magnitude ratio) were identified in many experiments, for example, thermal conductivity \cite{LuoTaillefer2009PRB},
London penetration depth \cite{Cho2014PRB,EvtushinskyBoriesenko2009NJP} and angle-resolved photoemission spectroscopy (ARPES) \cite{Ding2008EPL, EvtushinskyBoriesenko2009NJP,Nakayama2011PRB,Ota2014ARPES,LaserARPES2015}. In the heavily over-doped region, $x \geq 0.8$, a gap with line nodes was identified  by thermal
conductivity \cite{ReidTaillefer2012PRL, HongLi2015CPL_BaK122_heat_transport, WatanabeMatsuda2014PRB}, London penetration depth \cite{Hashimoto2010PRB_KFe2As2}
and ARPES \cite{Ding2008EPL,Ota2014ARPES}. Although some thermodynamic \cite{Hardy2014JPSJ,Hardy16} and small-angle neutron measurements \cite{SANS11} have reported tiny full gaps instead, there is a consensus that the gap anisotropy is very strong.

An important feature of the overdoped Ba122 system is the Lifshitz transition reported for both electron-doped \cite{LifshitzFeCo2010} and hole-doped \cite{Xu2013PRB} compounds. In the material of interest here, BaK122, there is a series of Lifshitz transitions in the $x\sim$0.7 to 0.9 region that result in the replacement of electron-like pockets at the $M$ point by hole-like pockets \cite{Xu2013PRB,Ding2015JPCM}, although a more precise determination of the critical compositions and the exact evolution of the three-dimensional band structure is still lacking. Indeed, this marked change in the electronic band structure must be taken into account when trying to explain the observed evolution of the superconducting properties with doping. One of the problems is the absence of systematic studies for a sufficient number of different compositions with reliably established values of \(x\), spanning the whole doping range. Here, we measured  16 different compositions  with $x$ values determined by the wavelength-dispersive spectroscopy (WDS) in each measured sample.

\begin{figure}[htb]
\includegraphics[width=8cm]{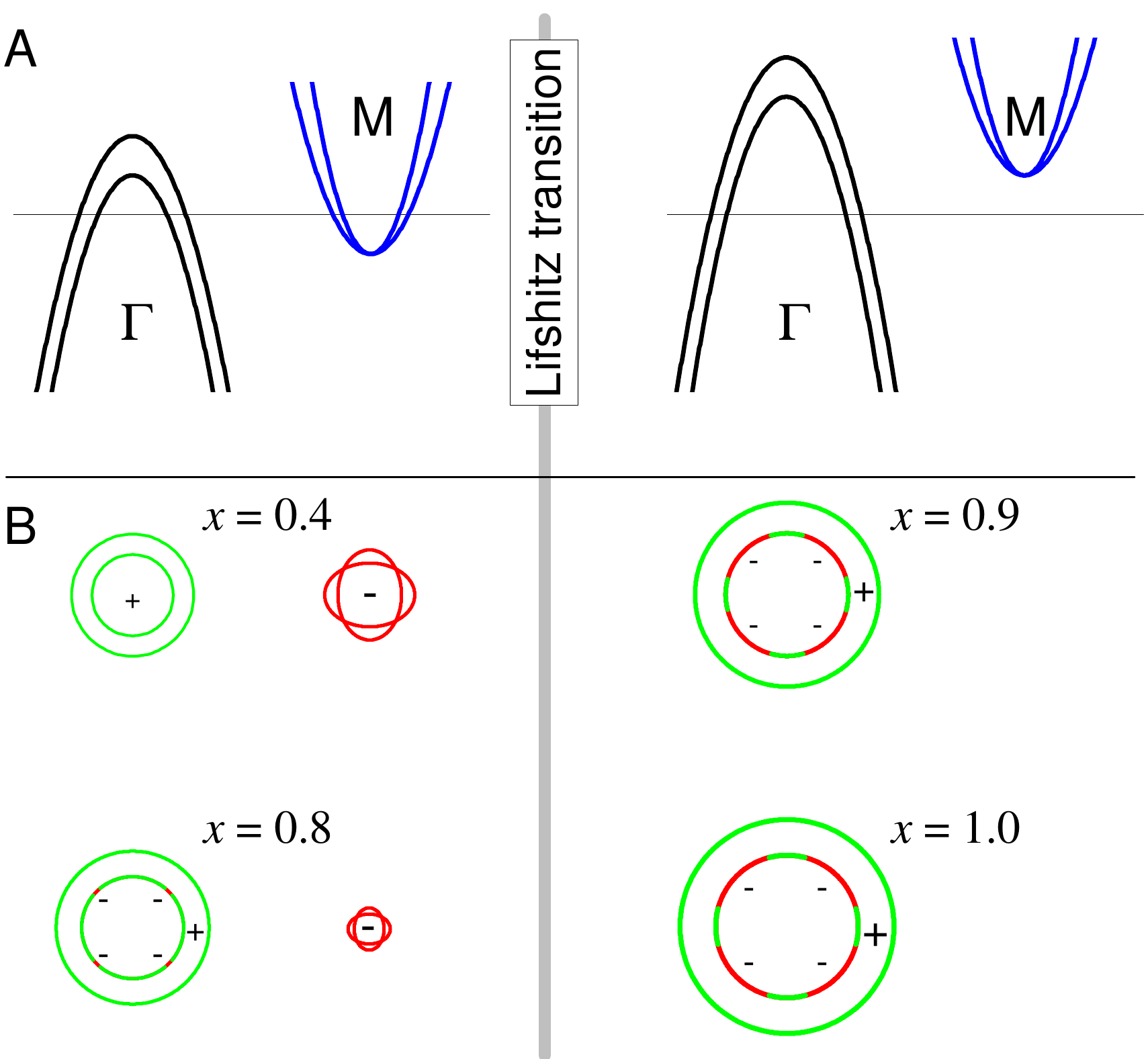}
\caption{(Color online) Schematic illustration of the effective band-structure and order parameter evolution with doping. (A) Change in the electronic band structure across the Lifshitz transition. The electron pocket at M is lifted but remains in the vicinity of $E_F$. The extended $s_{\pm}$ pairing survives but is shifted to the hole bands at the $\Gamma$ point. (B) Hole and electron pockets relevant for calculations with the sign-changing order parameter. Signs are encoded by green (+)  and red (-) colors.}
\label{fig1}
\end{figure}

Although there is an overall experimental consensus on the evolution with doping from large, isotropic to smaller,  highly anisotropic gaps  in BaK122, several possible theoretical interpretations exist. Most authors propose a crossover between two generalized $s-$wave states, where the usual configuration of isotropic gaps with opposite signs on the electron and hole pockets crosses over to a configuration with opposite signs on the hole pockets resulting in accidental nodes \cite{WatanabeMatsuda2014PRB}. This crossover may happen through an intermediate time-reversal symmetry broken $s+is$ state \cite{Maiti2015}. Some consider a transition from \spm~to $d-$wave either directly \cite{ThomaleBernevig2011PRL} or via an intermediate $s+id$ state \cite{Chubukov2012ARCMP, PlattWerner2012PRB, FernandesMillis2013PRL}. Still, others propose the existence of too-small-to-measure but finite ``Lilliputian" gaps \cite{Hardy2014JPSJ,Hardy16}. Because of the multitude of Fermi surface sheets and the absence of direct phase-sensitive experiments, it is difficult to pinpoint the most plausible explanation, and further studies are needed.

This is where the introduction of controlled artificial disorder becomes a very useful tool. In fact, impurity scattering is phase-sensitive and therefore can be used to at least narrow down possible scenarios. In most cases, only suppression of $T_c$ is studied, and even then, it can provide important information. For example, strong support for \spm~pairing was found in electron-irradiated Ba(Fe,Ru)$_2$As$_2$  \cite{Prozorov2014PRX}. Of course, in IBSs, it is rather difficult to draw definitive conclusions from $T_c$ suppression alone because of the many parameters involved in multiband pairbreaking \cite{Efremov2011}. Measurements of another disorder-sensitive parameter, for example low-temperature behavior of London penetration depth, can significantly constrain theoretical interpretations. This was suggested as a way to distinguish between \spm~and $s_{++}$ pairing \cite{WangHirschfeldMishra2013PRB}. This idea has been used to interpret the data in BaFe$_2$(As,P)$_2$ \cite{Mizukami2014NatureComm} and SrFe$_2$(As,P)$_2$ \cite{Strehlow2014}, where potential scattering lifted the nodes, thus proving them accidental and, therefore, lending a strong support to \spm~pairing.

Here, we measured low-temperature variation of London penetration depth, $\Delta \lambda(T)$, down to 50 mK in 16 different compositions of BaK122, for most of which the effect of artificial point-like disorder induced by 2.5 MeV electron irradiation at several doses was examined. By analyzing both the rate of $T_c$ suppression and changes in $\Delta \lambda (T) $, we conclude that increasing gap anisotropy on one of the hole bands at the $\Gamma$ point leads to the development of accidental nodes, and when the electron band no longer
 crosses the Fermi level at the $M$ point, \spm~pairing is realized between two hole bands. This is illustrated schematically in \fref{fig1}. In principle,  the incipient electron band can still play a role in interband interactions and pair-breaking scattering, but these effects are not qualitatively relevant here because superconductivity is supported by robust bands at the Fermi level \cite{Chen2015,Linscheid2016}.  We also discuss the possibility of a crossover from $s-$ to $d-$wave symmetry with increasing $x$ and
conclude that this is very unlikely, in line with ARPES studies that find accidental nodes on hole bands all the way up to $x=$1 \cite{OkazakiMatsuda2012Science,Ota2014ARPES}.

\begin{figure}[htb]
\includegraphics[width=8cm]{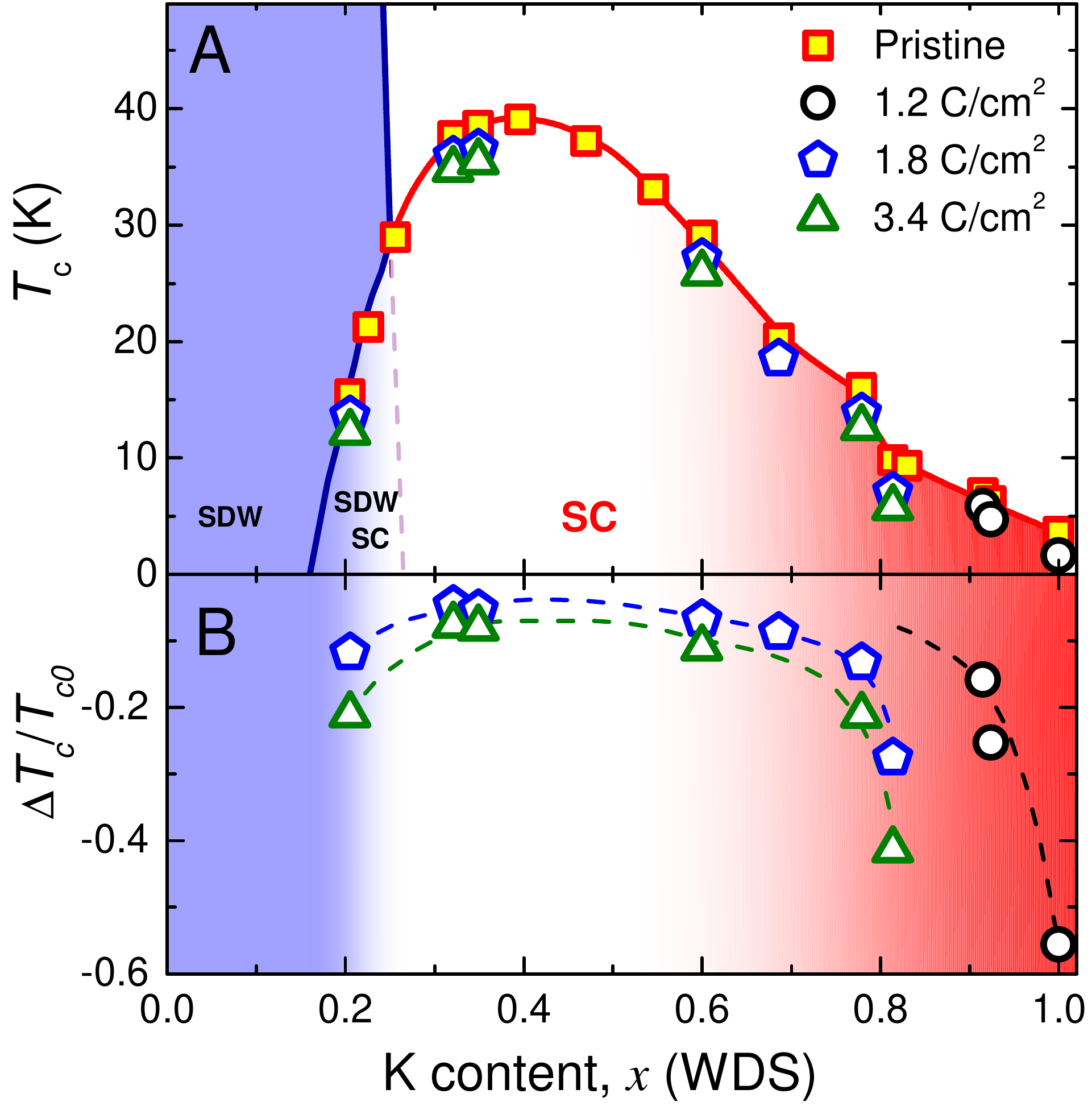}
\caption{(Color online) Temperature-composition phase diagram. (A) Composition-dependent superconducting transition temperature, $T_c(x)$, in pristine (squares) and electron-irradiated (other symbols, see legend) samples. SDW, spin-density wave; SC, superconducting phase. (B) Normalized $\Delta T_c / T_{c0}$. The largest $T_c$ suppression is found at $x \gtrsim 0.8$. The color shade indicates long-range magnetic order at small $x$ and crossover to nodal behavior at large $x$.}
\label{fig2}
\end{figure}

\section{Results}

\Fref{fig2}(A) shows the composition phase diagram of BaK122 compounds. The superconducting transition temperature, $T_c(x)$, was determined as the midpoint of the transition in penetration depth measurements (see \fref{fig1SI}). For pristine samples, $T_{c0}(x)$ shows its maximum value of 39 K at around $x \approx$ 0.40 and gradually decreases toward lower and higher $x$, forming a ubiquitous superconducting ``dome". Although the evolution of $T_c (x)$ is smooth in general, there is an apparent jump  near $x=$ 0.80. This anomaly correlates with the appearance of accidental nodes induced in this material as a consequence of the Lifshitz transition \cite{Maiti2012}. For most compositions shown in \fref{fig2}, the same samples were electron-irradiated and the London penetration depth was measured before and after each irradiation run. The relative change, $\left(T_c-T_{c0}\right)/T_{c0}$, is shown in \fref{fig2}B for the same samples as in \fref{fig2}A. The largest suppression of $\sim$47$\%$ per 1 C/cm$^2$ ($\sim$56.4$\%$ for 1.2 C/cm$^2$) was found in pure KFe$_2$As$_2$, and the smallest suppression was found in the optimally doped compounds.

\begin{figure}[tb]
\includegraphics[width=8cm]{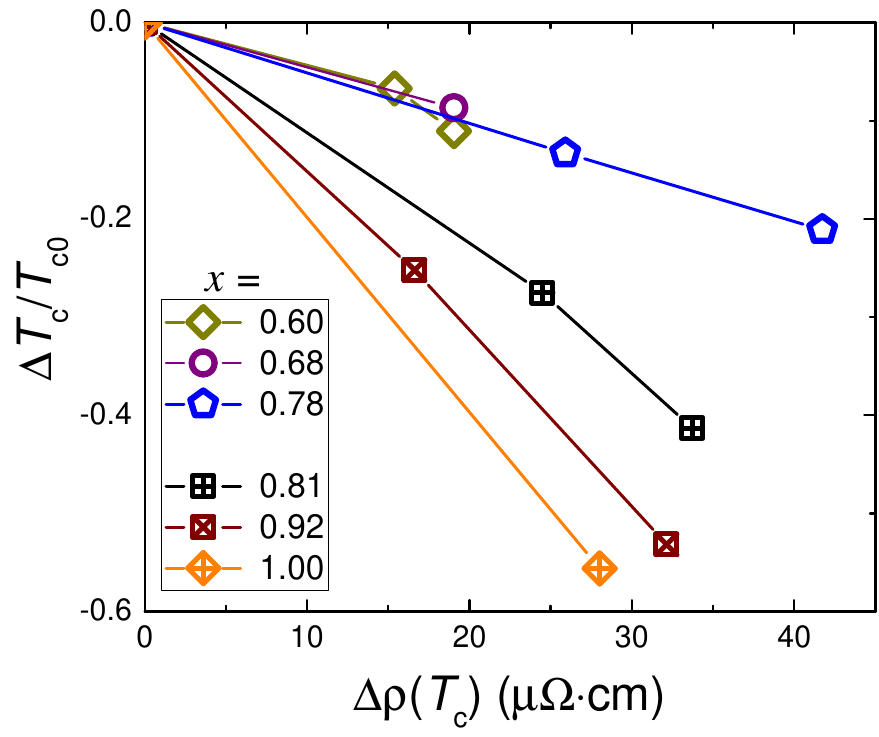}
\caption{(Color online) Normalized suppression, $\Delta T_c /T_{c0}$ versus resistivity at $T_c$ obtained from normal skin depth (see \fref{fig2SI}).}
\label{fig3}
\end{figure}

As shown in \fref{fig3}, the ``physically meaningful" normalized $T_c$ suppression plotted versus resistivity at $T_c$ shows a significant increase when transitioning from optimal to overdoped regimes. Note that because of magnetic ordering, these rates should not be compared directly to those of the underdoped regime, which require a separate analysis as a result of the competition between superconductivity and magnetism \cite{KimBaK122underdoped}.

In terms of the rate per irradiation fluence, the normalized suppression rate of optimally doped samples (\fref{fig4SI}(a)) is about 0.025 per 1 C/cm$^2$ and increases to 0.07 per 1C/cm$^2$ in the underdoped samples ($x=$0.22), consistent with our previous report \cite{Cho2014PRB}. In  sharp contrast, the suppression rate increases markedly in the far overdoped region, reaching 0.47 per 1 C/cm$^2$, which is 20 times larger than that of the optimally doped regime. All these numbers for the rate of $T_c$ suppression (i) are much greater than those expected from conventional $s_{++}$ pairing and (ii) can be explained within a generalized \spm~pairing model if one is allowed to tune gap anisotropy and the ratio of interband/intraband scattering [see Prozorov {\it et al.} \cite{Prozorov2014PRX} and references therein].

\begin{figure*}[tb]
\includegraphics[width=17cm]{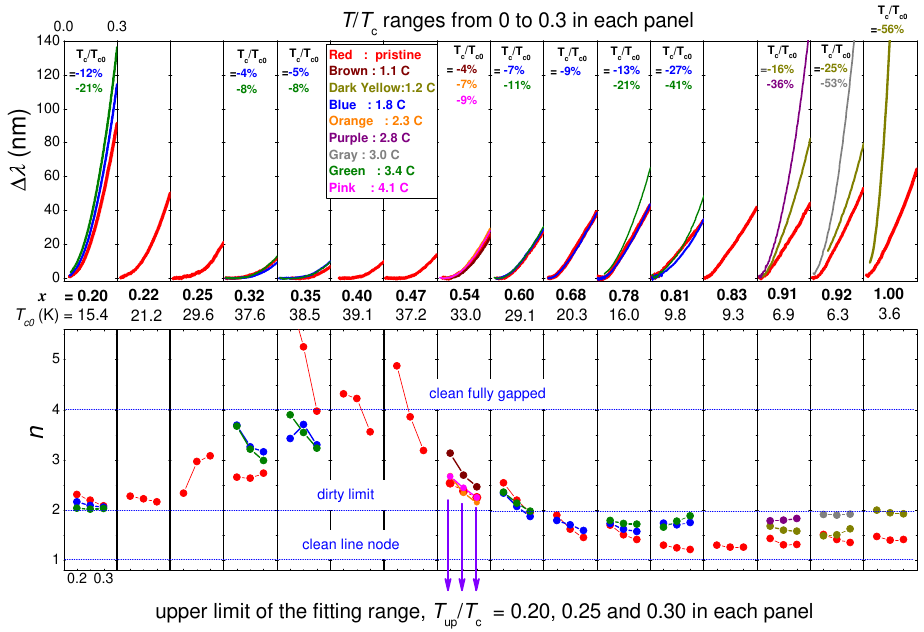}
\caption{(Color online) Evolution of temperature dependence of London penetration depth ($\Delta \lambda$). Upper panels: $\Delta \lambda (T/T_c )$ for 16 different compositions before and after electron irradiation. Each individual panel shows a low-temperature region of $T/T_c < 0.3$ (full-range curves are shown in \fref{fig1SI}). Lower panels: Exponent $n$ obtained from the power-law fitting, $\Delta \lambda = A(T/T_c)^n$. For each curve, three different upper-limit temperatures were used, $T_{up}/T_{c}$ = 0.20, 0.25 and 0.30, whereas the lower-limit was fixed by the lowest temperature.}
\label{fig4}
\end{figure*}

To understand the evolution of the superconducting gap with doping and disorder, we analyze the low-temperature behavior of the London penetration depth in terms of the power law, $\Delta \lambda (T) = A \left(T/T_c\right)^n$, as shown in \fref{fig4} and summarized in \fref{fig5}. To present the observed systematic trends, the upper panels of \fref{fig4} show $\Delta \lambda(T)$ on a fixed scale of 0 to 140 nm and at a temperature range of 0 to 0.3 $(T/T_c)$. (\fref{fig1SI} shows full-range curves). \Fref{fig5}A shows the composition variation of $\Delta \lambda (0.3T/T_c)$ reflecting the density of thermally excited quasi-particles. There is a clear trend of a marked increase in $\Delta \lambda$ as we move away from the optimal doping. At small $x$, this trend is naturally explained in terms of the competition between superconducting and SDW order \cite{PVCM2009,KimBaK122underdoped,Cho2014PRB}. The increase toward under-doped region is quite monotonic, whereas the increase toward $x=$1 is distinctly non-monotonic. There is even some decrease of $\Delta \lambda \left(0.3 T_c\right)$ around $x =$ 0.80, coincident with the anomaly in $T_c(x)$ (\fref{fig2}) and where the Liftshitz transition is believed to occur \cite{Xu2013PRB}. Similar non-monotonic behavior in the same region was reported before \cite{WatanabeMatsuda2014PRB}; thus, it seems that this is not an experimental abberation. In fact, this feature may signal the onset of accidental nodes near the Lifshitz transition \cite{Maiti2012}.

The lower panels of \fref{fig4} show the exponent, $n$, obtained in the power-law fitting. To examine the robustness of the power-law representation,  fitting of $\Delta \lambda (T/T_c)$ was performed from the base temperature up to three different upper limits, $T_{up}/T_c=$0.2, 0.25 and 0.30. The results are shown by three points in each frame of the lower panel of \fref{fig4}. \Fref{fig5}B summarized the composition and irradiation evolution of the exponent $n$ obtained at $T_{up}/T_c=$0.3. Horizontal lines show three principal limits of the exponent $n$ expected for different scenarios. A clean line nodal gap corresponds to $n=1$, whereas exponential behavior is experimentally not distinguishable from a large exponent ($n \geq$3 to 4). In all cases, $n=2$ is the terminal dirty limit for any scenario with pairbreaking (\spm~or $d-$wave), but it should be exponential for $s_{++}$ pairing where non-magnetic scattering is not pair-breaking.

At small $x$, in the coexistence regime, the gap anisotropy increases, but we find no evidence of nodes, consistent with the previous studies \cite{Cho2014PRB,KimBaK122underdoped}. This result argues against an $s_{++}$ gap structure, in which the reconstruction of Fermi surfaces due to SDW order must lead to robust nodes \cite{PVCM2009}. Upon electron irradiation, $T_c$ slowly decreases, suggesting moderate gap anisotropy and the presence of small but significant interband impurity scattering \cite{SDWSCVM2015}. Close to the optimal composition of $x=$0.4, the penetration depth exponent $n$ decreases significantly with irradiation, providing strong evidence for \spm~pairing. On the other hand, even upon a high-dose irradiation of 3.4 C/cm$^2$, the exponent remains greater than $n=$3, whereas $T_c$ decreases by 8\%, which is suggestive of robust full gaps. In a fully gapped $s_{++}$ state, the only effect of  disorder is to average the gap over the Fermi surface, leading inevitably to the increase of the minimum gap and therefore an increase in the exponent $n$ with disorder, contrary to our observations.

Moving to higher $x$ away from optimal composition, the gap anisotropy increases and the exponent $n$ for the pristine samples decreases. Upon irradiation, the gap anisotropy is smeared out and the exponent increases even in the \spm~case, provided that all bands are still fully gapped and the intraband impurity scattering is dominant. This is apparently the case for $x=0.54$. For yet higher doping levels, the anisotropy becomes so strong that the system develops accidental nodes ($n\rightarrow 1$), which, in this case, are apparently not lifted by the disorder \cite{Mishra2009PRB}. This is possibly due to (i) the substantial change in the electronic band structure approaching the Lifshitz transition and/or (ii) substantial interband impurity scattering. Note that this evolution is very different from the isovalently substituted BaFe$_2$(As,P)$_2$ \cite{Mizukami2014NatureComm} in which line nodes are found at all $x$ values and the band structure is unchanged.
In our case, at large $x$, the exponential temperature dependence in pristine samples changes to $\sim T^2$ at around $x$ = 0.60 and tends toward $\sim T$ at $x \geq$ 0.80, indicative of gaps with line nodes. Unlike the optimally doped region, the electron irradiation is much more effective in decreasing $T_c$, that is, 41\% upon 3.4 C/cm$^2$ ($x =$ 0.81) and 56\% upon 1.2 C/cm$^2$ ($x =$ 1.00). Nevertheless, the exponent $n$ never exceeds 2.

\section{\spm~superconductivity in the overdoped region}

In previous studies of London penetration depth and thermal conductivity in overdoped samples, nodal behavior was
identified above $x=0.8$ and attributed to a crossover from a fully isotropic $s-$wave state with sign change
between electron and hole pockets to a new type of \spm~pairing with sign change between
the hole pockets \cite{WatanabeMatsuda2014PRB}, which also acquired accidental nodes. Thermal conductivity measurements of the end member, at $x=$1, indicate line nodes and were interpreted in terms of $d-$wave pairing \cite{ReidProzorovTaillefer2012SST,ReidTaillefer2012PRL}, which was also claimed theoretically \cite{ThomaleBernevig2011PRL}. Here, we add an additional restriction on possible interpretations by looking simultaneously at the variation of $T_c$ and temperature-dependent London penetration depth with controlled point-like disorder. As we mentioned above, the suppression of $T_c$ with irradiation at optimal doping rules out a global $s_{++}$ state. Now, the challenge is to begin with a ``conventional" (nodeless) \spm~state and determine whether a reasonable model of superconducting gap and its evolution with composition can be constructed to describe all experimental results. We find that a generalized sign-changing \spm~state with accidental nodes can be used to describe the entire phase diagram, including a crossover from nodeless to nodal gap. The novel assertion of our approach is that with electron pockets absent above Lifshitz transition, $x>0.8$, the \spm~pairing shifts to hole pockets, naturally resulting in a nodal state.

We use the self-consistent $t-$matrix formalism and sign-changing \spm~state to describe both the London penetration depth and $T_c$ suppression rate for different levels of disorder. To keep the analysis tractable and to fit the experimental data, we minimize our parameter set by working in the 2Fe-BZ and modeling the gap structure as shown schematically in \fref{fig1}. Specifically, before the Lifshitz transitions, two electron pockets at the $M$ point and two hole pockets at the $\Gamma$ point are each modeled as a single $C_4$ symmetric pocket with gap, $\Delta_{e}(\phi) = \Delta_{e}^{iso} + \Delta_{e}^{ani} \cos 4\phi$ and $\Delta_{h_1}(\phi) = \Delta_{h_1}^{iso} + \Delta_{h_1}^{ani} \cos 4\phi$, respectively. Here, angle $\phi$ is measured from the zone diagonal. After the Lifshitz transition, the electron pockets disappear, and the two model bands now correspond to two hole pockets.  Each hole pocket gap is  now modeled independently, with its own isotropic and anisotropic components, $\Delta_{h_2}=\Delta_{h_2}^{iso} + \Delta_{h_2}^{ani} \cos 4\phi$. We realize that the actual band structure is more complex, and its evolution across the Lifshitz transition involves several bands changing across the Brillouin zone. However, we find that a model with two effective gaps each having isotropic and anisotropic parts is sufficient to explain the observed results.

First, we fit the data for the pristine samples and then include impurity scattering within self-consistent
$t$-matrix formalism \cite{Hirschfeld1993PRB,Hirschfeld88,MGH2011,MVHV2009}. We model the defects induced by electron irradiation as point-like scatterers, which scatter between the bands with a certain (interband) amplitude  and within the same band with another (inband) amplitude. The presence of interband impurity scattering and the relative sign change between these two bands are necessary to explain the $T_c$ suppression and penetration depth in the irradiated samples (see the Supplementary Materials for details of the fitting procedure). The obtained gap amplitudes are shown in \fref{fig6} as a function of composition, $x$. It is important that the average gap $h_1$ on one of the hole bands changes its sign with increasing $x$. This is essential to fit the $T_c$ suppression and penetration depth on equal footings in a self-consistent manner  (see \fref{fig4SI}(b)).
Without a relative sign change between the hole pockets, even a strong interband impurity scattering will average the gap anisotropy, leading to a weak suppression of $T_c$ and activated behavior in the temperature dependence of the low-temperature penetration depth, which is not in agreement with the data. The obtained evolution of gaps suggests a new paradigm where an \spm~superconducting state with relative sign change between the hole and electron pockets at moderate doping levels evolves into a \spm~state with the sign change between the hole bands with accidental line nodes. This evolution of the gap structure is shown schematically in \fref{fig1} and is the central result of this paper.

\begin{figure}[tb]
\includegraphics[width=8cm]{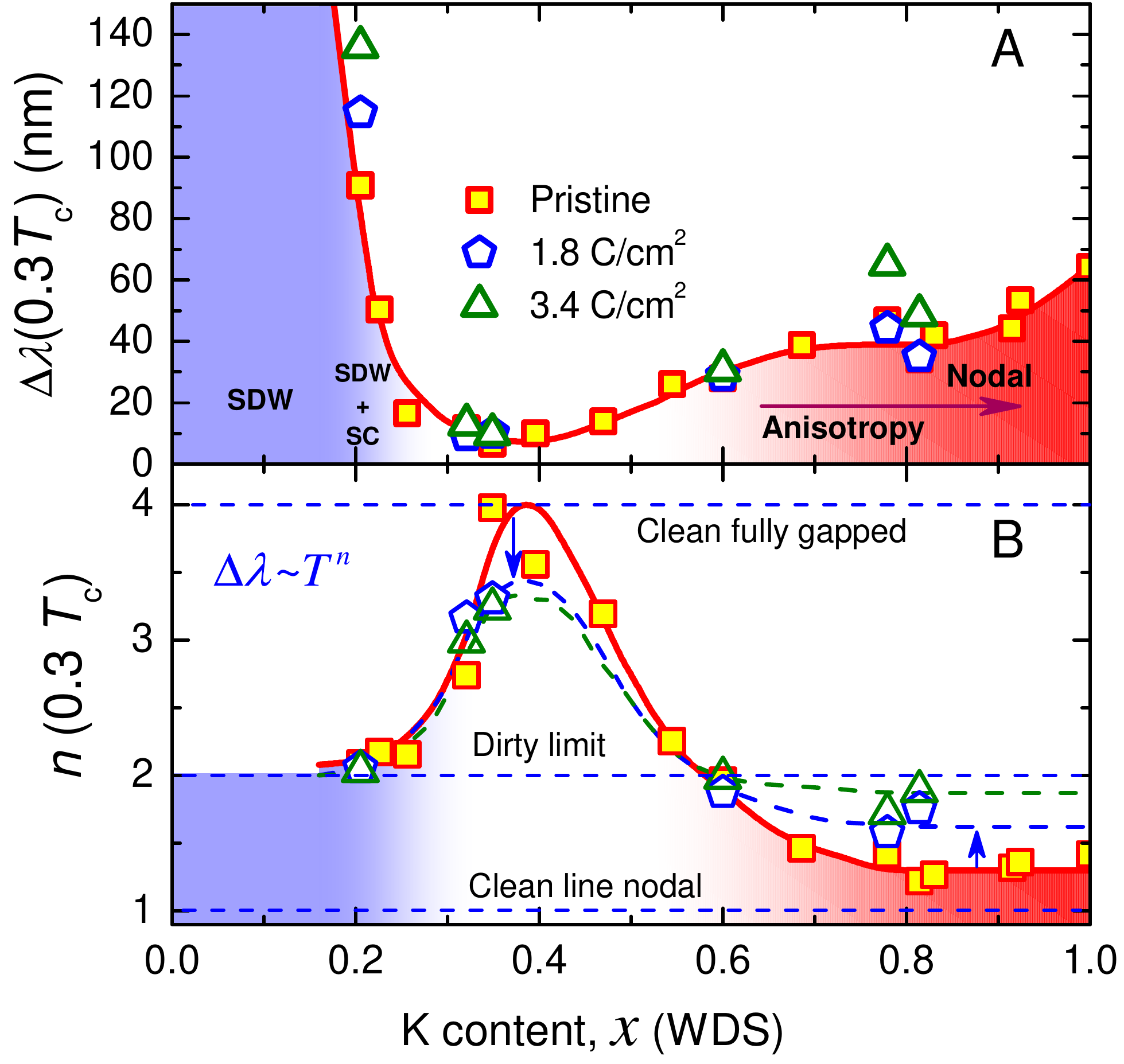}
\caption{(Color online) Absolute change of $\Delta \lambda$ from 0 to 0.3 $T_c$ for all compositions. (A) The change in the London penetration depth, $\Delta \lambda (0.3 T_{c})$, versus $x$ for pristine and post-irradiated samples. (B) Composition dependence of the power-law exponent $n$ for pristine and irradiated samples. As the irradiation dose increases, the exponent approaches, but never exceeds, the value of $n=$2. }
\label{fig5}
\end{figure}

\begin{figure}[htb]
\includegraphics[width=8cm]{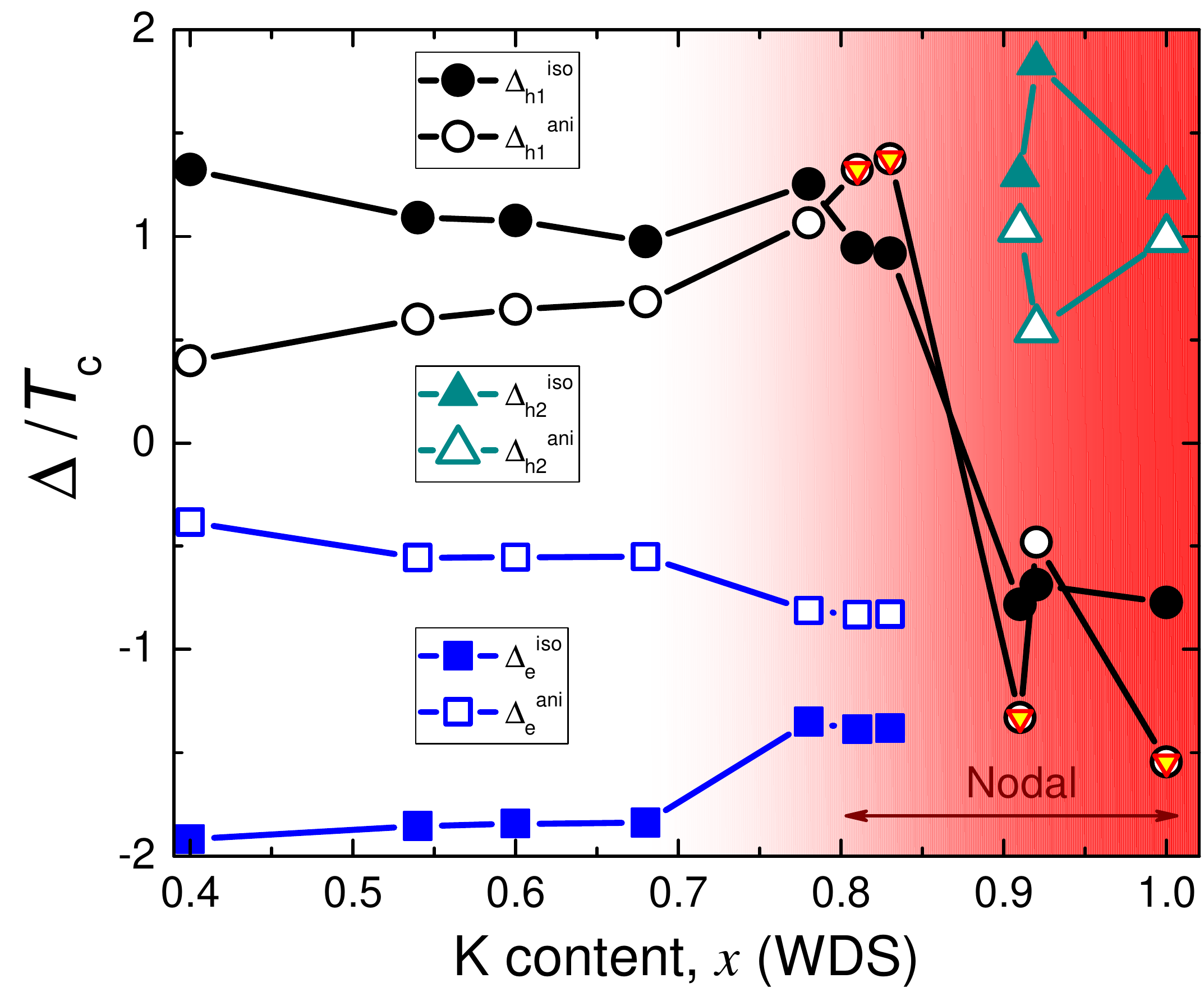}
\caption{(Color online) Evolution of the superconducting gaps in BaK122 with composition, $x$, obtained from self-consistent $t-$matrix
fitting (see \fref{fig3SI}) as described in the text. Assumed angular variations of the gaps is shown schematically in \fref{fig1}.
As long as the isotropic part is greater than the anisotropic one, the state is nodeless (that is, for $x<$0.8). In the opposite limit, the nodes appear. This is shown by inscribed triangles in the figure for $h_1$ contribution. Consequently, the \spm~pairing switches from hole-electron pockets below the Lifshitz transition to hole-hole above.}
\label{fig6}
\end{figure}

\begin{figure}[htb]
\includegraphics[width=1\linewidth]{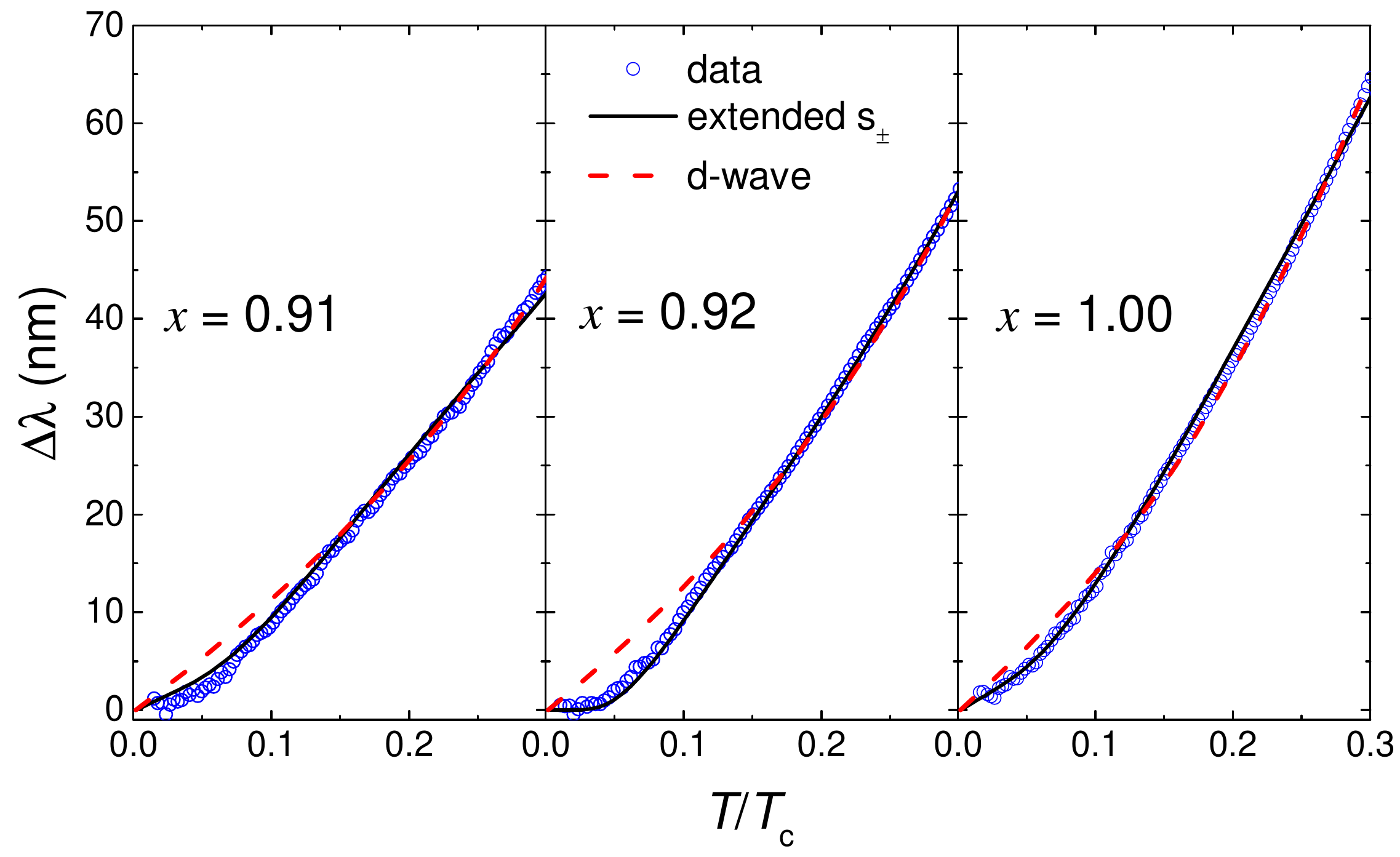}
\caption{(Color online) The change in penetration depth for the $x > 0.9$ samples fitted with symmetry-imposed $d-$wave and \spm~state. For $d-$wave fit, both the hole bands are assumed to have gaps of $\Delta_{h1/h2} \cos 2\phi$ form. The gap magnitudes ($\Delta_{01},\Delta_{02}$) for dopings x=0.91,0.92 and 1.00 are
(1.5,1.8), (1.6,1.2) and (1.0,1.2) respectively in units of $T_c$.}
\label{fig7}
\end{figure}

We note that if one concentrates exclusively on Fermi surface integrated quantities, such as thermal conductivity or penetration depth, distinguishing $d-$wave states from anisotropic, deeply nodal $s-$wave states can be very difficult. As shown in  \fref{fig7}, both $d-$wave and anisotropic $s_\pm$ states give reasonable fits to the pristine penetration depth data for $x=1.0$.
Furthermore, distinguishing on the basis of disorder is difficult because here we do not have a well-established link between the pair-breaking rate and irradiation dosage; thus, it is possible  to find parameters for either ``dirty $d-$wave" or ``dirty nodal $s-$wave" cases that fit both the $\Delta\lambda$ and $\Delta T_c$ data for the single $x=1$ sample. However,  on the basis of  the fits to the heavily K-doped alloys near $x=0.9$  in Fig.{\ref{fig7}}, we see that there is substantial additional curvature at low temperatures that is incompatible with the $\cos2\phi$ $d-$wave. It is conceivable that a strong anti-phase $\cos6\phi$ component in a $d-$wave state could fit the penetration depth data. However, there is no theory in support of such a state, and we therefore conclude that the superconducting condensate in this system has $s-$wave symmetry throughout the phase diagram and simply evolves in an anisotropic manner as roughly depicted in \fref{fig1}.
In \fref{fig7}, we show a comparison between the state with accidental nodes and a $d-$wave state for $x=0.91$ and $x=0.92$. For $x=0.91$ and $0.92$, we see the incompatibility of a $d-$wave gap with experimental data. However, for $x=1.0$, both $d-$wave and $s_\pm$ state with accidental nodes can fit the data. Thus, we cannot rule out a crossover between $s-$wave and $d-$wave symmetries between  $x=0.92$ and $x=1.0$. However, ARPES measurements provide a strong argument against this scenario \cite{Ota2014ARPES}.

An additional argument favoring \spm~pairing with accidental nodes over the $d-$wave  is the non-monotonicity of $\Delta\lambda(0.3T_c,x)$ near $x\sim$0.8 (see \fref{fig5}). Because of an overall decrease in $T_c$ in the overdoped region, the normal fluid density in an isotropic $s-$ or $d-$wave state is expected to monotonically increase. On the other hand, this non-monotonicity occurs naturally during a smooth onset of accidental nodes -- where a fully gapped $\Delta(\phi)$  near the expected nodal region transits to a linear-in-$\phi$ dependence through an intermediate quadratic, $\Delta(\phi) \sim\phi^2$, dependence. Accidental nodes not only are more probable for $s-$wave pairing as opposed to $d-$wave pairing but also are expected to appear around the Lifshitz transition \cite{Maiti2012}. Of course, non-monotonicity of $\Delta \lambda (0.3T_c,x)$ does not uniquely imply accidental nodes, but accidental nodes naturally lead to the observed non-monotonic behavior. This scenario can also explain variations observed at the lowest temperatures for close compositions, such as $x=$ 0.91 and 0.92 (see \fref{fig7}).

We emphasize that our analysis of the rate of $T_c$ suppression by non-magnetic scattering supports accidental nodes in an $s_{\pm}$ state rather than in an $s_{++}$ state.  Although a small gap could be present in the $x=0.92$ sample, at $x=1$, our data and fitting appear to rule this out. However, within our experimental temperature range, down to 50 mK, it is impossible from the penetration depth alone to definitively rule out gaps on the order of 0.1 meV or smaller, such as those suggested by the analysis of the specific heat experiments \cite{Hardy2014JPSJ,Hardy16}.

Nevertheless, our systematic measurements and analysis of many different compositions add to the growing experimental support for the $s-$wave  origin of the pairing interaction near $x\sim 1$ (and therefore over the whole phase diagram). This in turn indicates that any competing $d-$wave channel, as predicted by many theoretical approaches, is competitive but sub-leading all the way up to $x=1$. The extent of this competition can be addressed by probing collective modes in the non-pairing channel using other experimental techniques (for example, Raman Scattering). We note that some studies of the $x=1$ composition under pressure also propose a change of pairing symmetry from $d-$-wave to $s-$wave \cite{Tafti2015}. Our work poses severe difficulties for such an interpretation.

\section{Conclusions}
In conclusion, we used deliberately introduced point-like disorder as a phase-sensitive tool to study the compositional evolution of the superconducting gap structure in BaK122. Measurements of both the low-temperature variation of London penetration depth and $T_c$ suppression provided stringent constraints on the possible gap structures. By using a generalized \spm~model and $t-$matrix calculations, we were able to describe the compositional evolution of the superconducting gap, including a crossover from nodeless to nodal concomitant with the Lifshitz transition. Our model provides a natural interpretation of the rich physics of this system and shows that \spm~pairing is a very robust state of iron pnictides.

\section{Materials and Methods}
\subsection{Crystal growth}
We developed an inverted temperature gradient method to grow large and high-quality single crystals of BaK122. The starting materials--Ba and K lumps, and Fe and As powders--were weighed and loaded into an alumina crucible in a glove box. The alumina crucible was sealed in a tantalum tube by arc welding. The tantalum tube was then sealed in a quartz ampoule to prevent the tantalum tube from oxidizing in the furnace.  The crystallization processes from the top of a liquid melt help to expel impurity phases during the crystal growth, compared to the growth inside the flux. Details of the growth and detailed characterization for the entire dome can be found elsewhere \cite{LiuProzorovLograsso2014PRB, Liu2013PRB}.

\subsection{Sample characterization and selection}
Sixteen different compositions ranging from $x =$ 0.20 to 1.00 were identified using WDS. More than one sample of each composition was studied. The crystals had typical dimensions of 0.5 mm $\times$ 0.5 mm $\times$ $0.03$ mm.

All samples were pre-screened using a dipper version of the tunnel diode resonator (TDR) technique \cite{NaFeCo2012transport}, using the sharpness of the superconducting transition as a measure of quality in each particular piece. After this pre-screening, chemical composition of each individual sample was determined using WDS in a JEOL JXA-8200 electron microprobe. In each sample, the composition was measured for 12 points per surface area and averaged \cite{LiuProzorovLograsso2014PRB}.

The variation of in-plane London penetration depth $\Delta \lambda (T)$ was measured
down to 50 mK using a self-oscillating TDR \cite{Prozorov2000PRB,Prozorov2006SST,ProzorovKogan2011RPP}. To study the effect of disorder, $\Delta \lambda (T)$ for each crystal was measured before and after the irradiation.

\subsection{Electron Irradiation}
Irradiation by 2.5 MeV electrons was performed at the SIRIUS Pelletron in Laboratoire des Solides Irradi$\acute{e}$s at $\acute{E}$cole Polytechnique (Palaiseau, France). The electrons created Frenkel pairs that acted as point-like atomic defects. Throughout the paper, the total acquired irradiation dose was conveniently measured in coulombs per squre centimeter, where 1 C/cm$^2 = $6.24 $\times$ 10$^{18}$ electrons/cm$^2$. With a calculated  head-on collision displacement energy for Fe ions of 22 eV and cross section to create Frenkel pairs at 2.5 MeV of 115 barn (b), a dose of 1 C/cm$^2$ resulted in about 0.07\% of the defects per iron site. Similar numbers were obtained for other sites - cross sections for Ba and As are 105 and 35 b, respectively. It is known that the interstitials migrate to various sinks (surface, dislocations, etc) and vacancies remain in the lattice. The electron irradiation was conducted in liquid hydrogen at 22 K, and recombination of the vacancy-interstitial pairs upon warming up to room temperature was 20 to 30 \%, as measured directly from the decrease of residual resistivity \cite{Prozorov2014PRX}. After initial annealing, the defects remain stable, which was established from re-measurements performed several months (up to more than a year) apart. In addition, by measuring the Hall coefficient, it was determined that electron irradiation did not change the effective doping level; neither did it induce a measurable magnetic signal, which would be detected in our sensitive TDR measurements.

\section*{Acknowledgements}
We thank A. Chubukov, R. Fernandes, Y. Matsuda, I. Mazin, T. Shibauchi and L. Taillefer for useful discussions.

\emph{Funding}: This work was supported by the U.S. Department of Energy (DOE), Office of Science, Basic Energy Sciences, Materials Science and Engineering Division. Ames Laboratory is operated for the U.S. DOE by Iowa State University under contract DE-AC02-07CH11358. We thank the SIRIUS team, O. Cavani, B. Boizot, V. Metayer, and J. Losco for running electron irradiation at $\acute{E}$cole Polytechnique [supported by EMIR (R$\acute{e}$seau national d'acc$\acute{e}$l$\acute{e}$rateurs pour les Etudes des Mat$\acute{e}$riaux sous Irradiation) network, proposal 11-11-0121]. V.M. acknowledges the support from the Laboratory Directed Research and Development Program of Oak Ridge National Laboratory, managed by UT-Battelle, LLC, for the U. S. DOE. P.J.H. and S.M. were partially supported by NSF-DMR-1005625.

\emph{Authour contributions}: K.C. and S.T. conducted London penetration depth measurements. K.C. processed and analyzed raw data. M.K. led electron irradiation work. M.K., R.P. and K.C. performed electron irradiation. M.A.T. handled all sample preparation and transport measurements. Y.L. and T.A.L grew single crystals. W.E.S. performed WDS measurements. V.M., S.M., and P.J.H. led theoretical work. V.M. performed $t-$matrix fitting and data analysis. R.P. devised and coordinated the project. All authors extensively discussed the results, the models and the interpretations. All authors contributed to writing the manuscript.

\emph{Competing Interests}: The authors declare that they have no competing interests.

\emph{Data and materials availability}: All data needed to evaluate the conclusions in the paper are present in the paper and/or the Supplementary Materials. Additional data related to this paper may be requested from the authors.

\bibliographystyle{ScienceAdvances}

\clearpage
\newpage

\section*{Supplementary Materials}
\setcounter{figure}{0}
\setcounter{equation}{0}

\makeatletter
\renewcommand{\thefigure}{S\@arabic\c@figure}
\makeatother

\makeatletter
\renewcommand{\theequation}{S\@arabic\c@equation}
\makeatother
\onecolumngrid

\section{London penetration depth}
\begin{figure*}[htb]
\includegraphics[width=14cm]{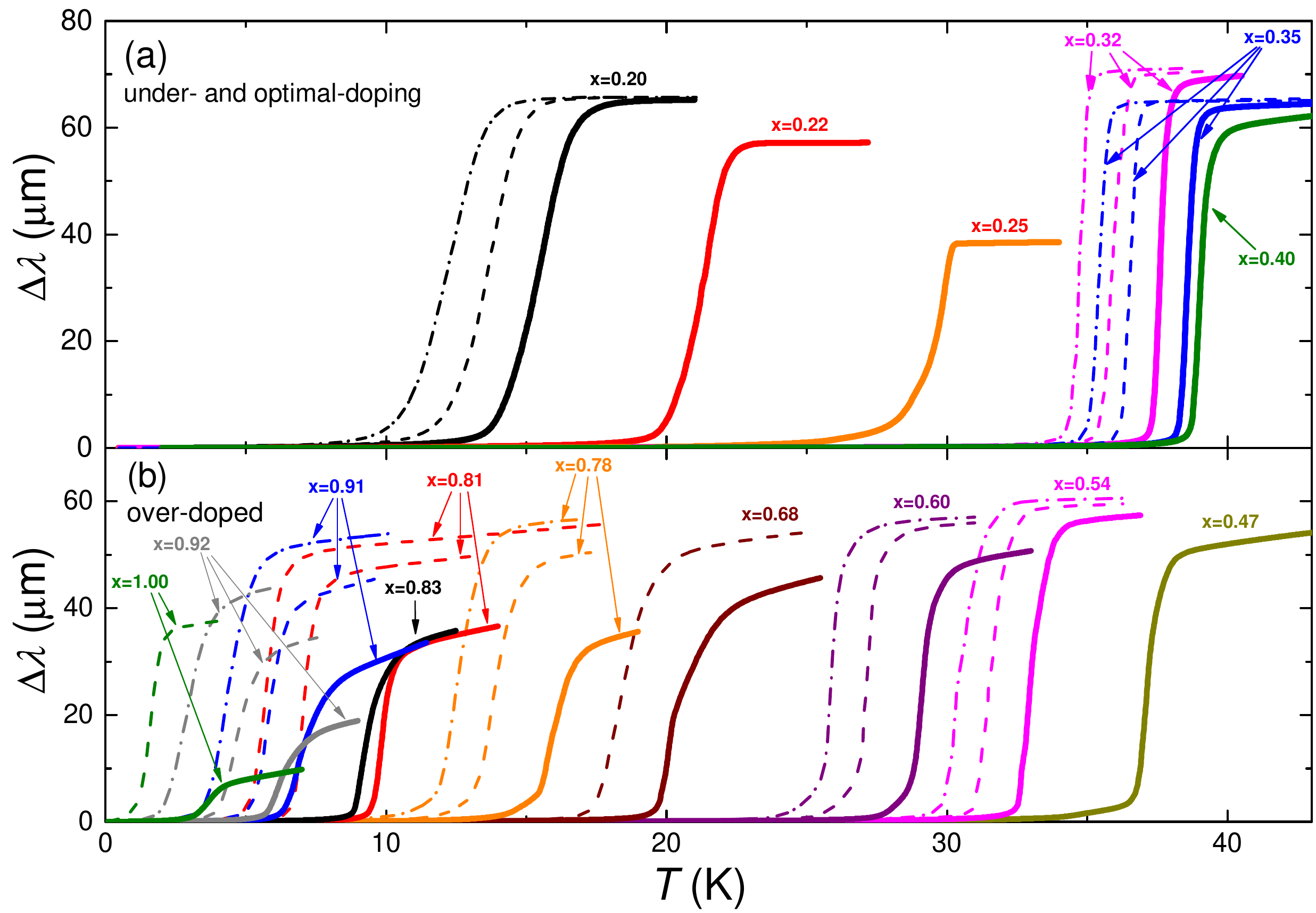}
\caption{(Color online) Full transition curves of $\Delta \lambda (T)$ for the
studied samples. Panel (a) shows under- and optimal- doping region. Panel (b) shows
overdoped regime. Solid lines are for pristine samples, dashed lines are for irradiated
as indicated in legend. Irradiation doses are the same as in \fref{fig4}.}
\label{fig1SI}
\end{figure*}

\begin{figure}[htb]
\includegraphics[width=14cm]{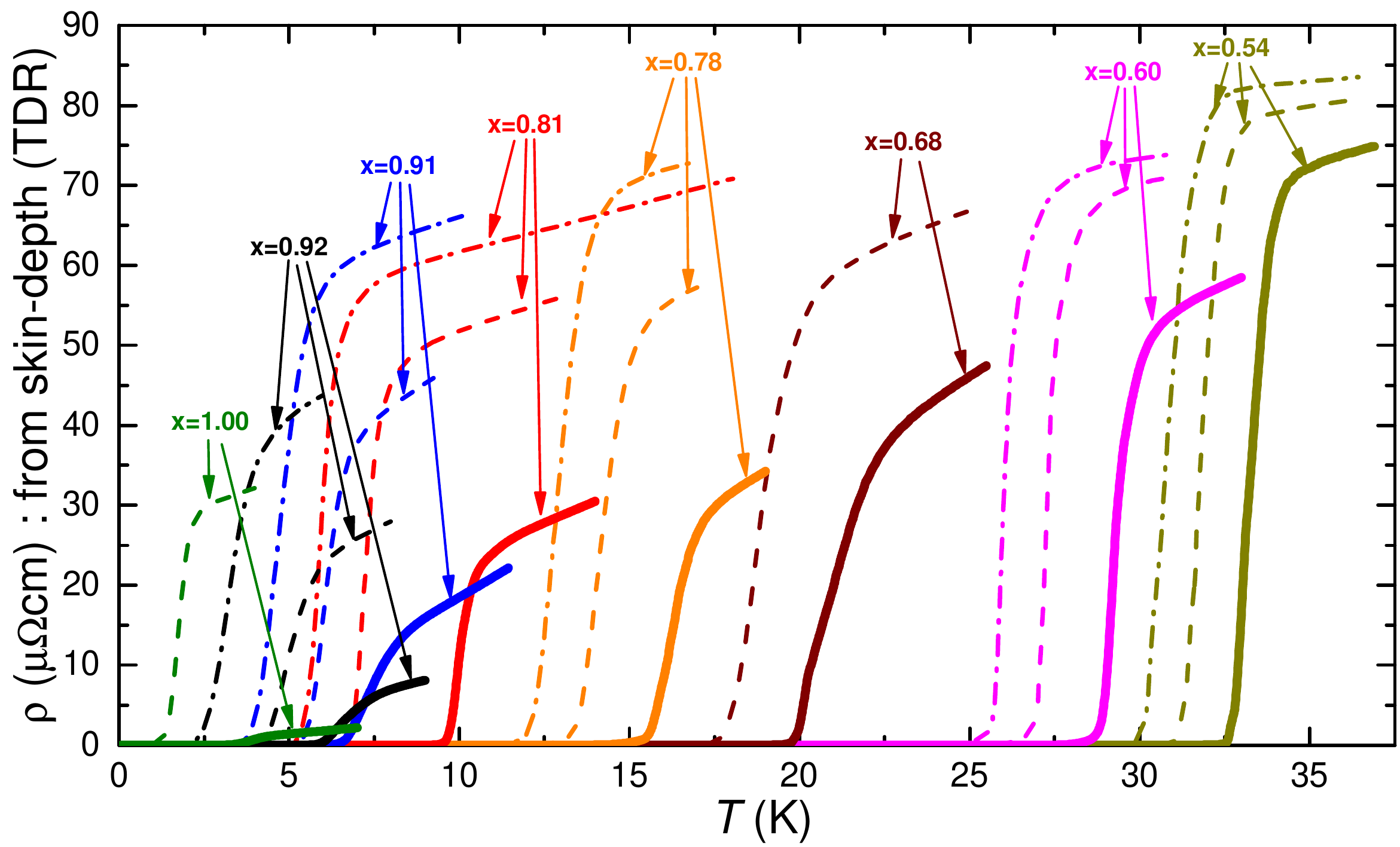}
\caption{(Color online) Resistivity estimated from the skin depth (TDR).}
\label{fig2SI}
\end{figure}

\Fref{fig1SI} shows London penetration depth $\Delta \lambda (T)$ in the full temperature range for all compositions before (solid lines) and after (dashed lines) electron irradiation. In our measurements, the saturation above $T_c$ occurs when $\lambda(T)$ becomes of the order of the sample size (size-limited) or normal-state skin depth. In the former (size-limited) case, the curve is flat above $T_c$ which is the case for $x =$ 0.20 - 0.40. However, in the latter (skin-depth limited) case, the curve shows temperature dependence above $T_c$ since the skin depth changes with temperature which is the case for $x =$ 0.54 - 1.00. Then, the resistivity, $\rho(T)$, can be evaluated from the measured skin depth, $\delta(T)=(\rho(T) c^2/2 \pi \omega)^2$, where $\omega = 2 \pi f$ is the resonator frequency. In \fref{fig2SI}, the resistivities obtained from the skin-depth are plotted. In general, the resistivity at $T_c$ of \bak~ decreases monotonously with the increase of $x$ \cite{LiuProzorovLograsso2014PRB}. The resistivity change upon irradiation at $T = T_c$ ($\Delta \rho (T = T_c)$), obtained from the skin depth, is plotted in comparison with $\Delta T_c / T_{c0}$ in \Fref{fig3}.

\section{\emph{t}-matrix fitting procedure}
A minimal two band model is used to fit the penetration depth data for pristine samples. One of the band represents the hole pocket throughout the phase diagram. Before the Lifshitz transition, the second band represents the electron band and after the
Lifshitz transition the second band represents the other hole band. We use gap magnitudes as the fitting parameters and along with an overall scaling factor, which takes care of the Fermi velocities and density of states for various doping. Model gap functions are

\begin{eqnarray}
 \Delta_1 &=& \Delta_{01} \left( 1.0 + r_1 \cos 4\phi \right) \\
 \Delta_2 &=& \Delta_{02} \left( 1.0 + r_2 \cos 4\phi \right).
 \end{eqnarray}

\begin{figure*}[htb]
\includegraphics[width=14cm]{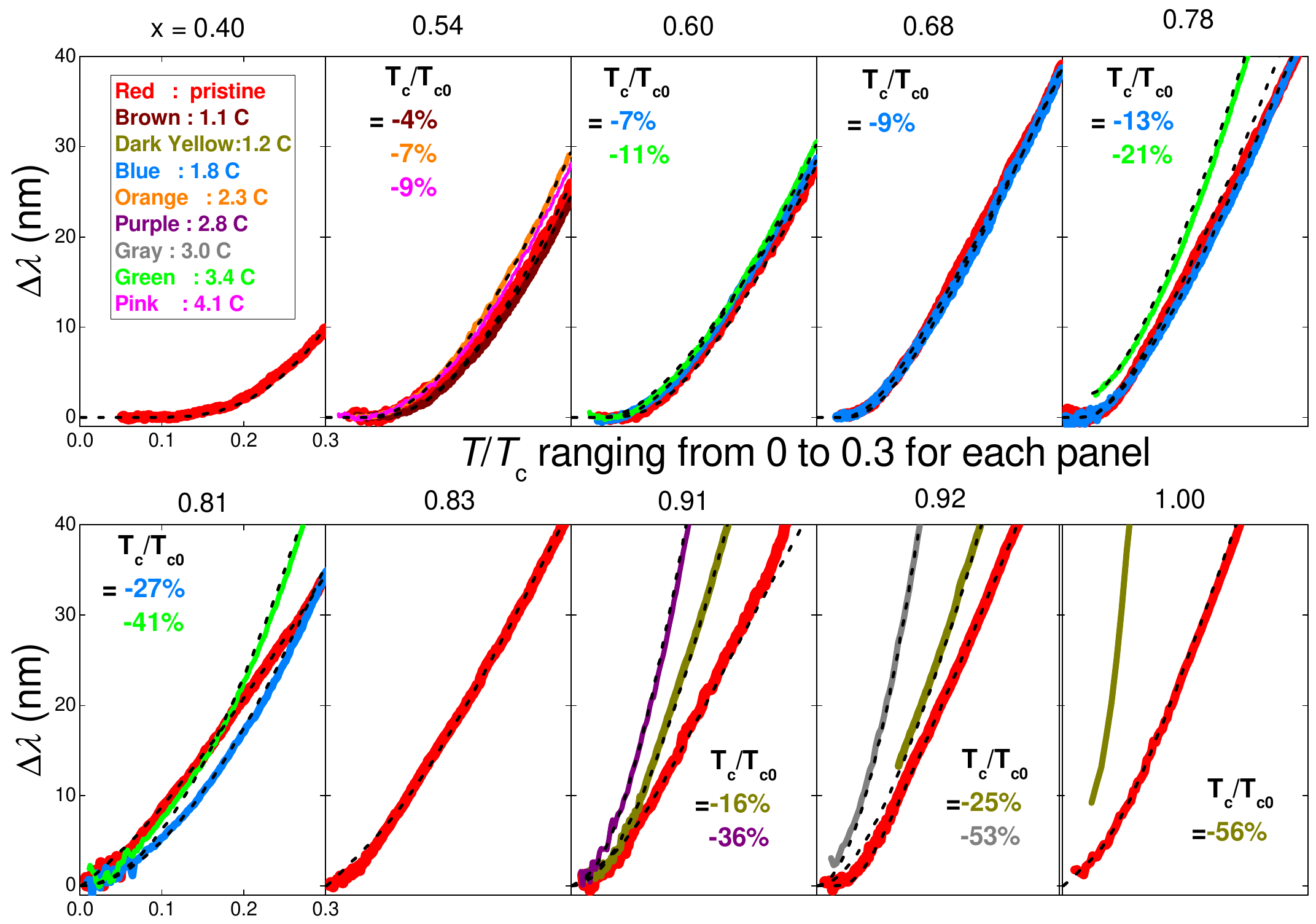}
\caption{(Color online) $t-$matrix fitting of the London penetration depth for
compositions spanning the superconductivity dome. The extracted gap magnitudes are plotted in \fref{fig6}.}
\label{fig3SI}
\end{figure*}

We first fit the low temperature penetration depth for the pristine samples, and find the gap values in the units of pristine sample's transition temperature $T_{c0}$. The fits are shown in \fref{fig3SI}.

Once we determine the gaps, we find the interactions which generate these gaps within the weak-coupling BCS approximation. We parameterize the interaction potential in a simple form, where, to reduce the number of parameters, we have assumed the angular form factors in the interactions to be the same as in the one in the gap structure

\begin{eqnarray}
 V_{11} &=&  V_1 \left( 1.0 + r_1 \cos 4\phi \right)  \left( 1.0 + r_1 \cos 4\phi^\prime \right)\\
 V_{12} &=& V^\prime \left[ \left( 1.0 + r_1 \cos 4\phi \right)  \left( 1.0 + r_2 \cos 4\phi^\prime \right)+\left( 1.0 + r_2 \cos 4\phi \right)  \left( 1.0 + r_1 \cos 4\phi^\prime \right)  \right] \\
  V_{22} &=&  V_2 \left( 1.0 + r_2 \cos 4\phi \right)  \left( 1.0 + r_2 \cos 4\phi^\prime \right).
\end{eqnarray}

\begin{figure}[htb]
\includegraphics[width=14cm]{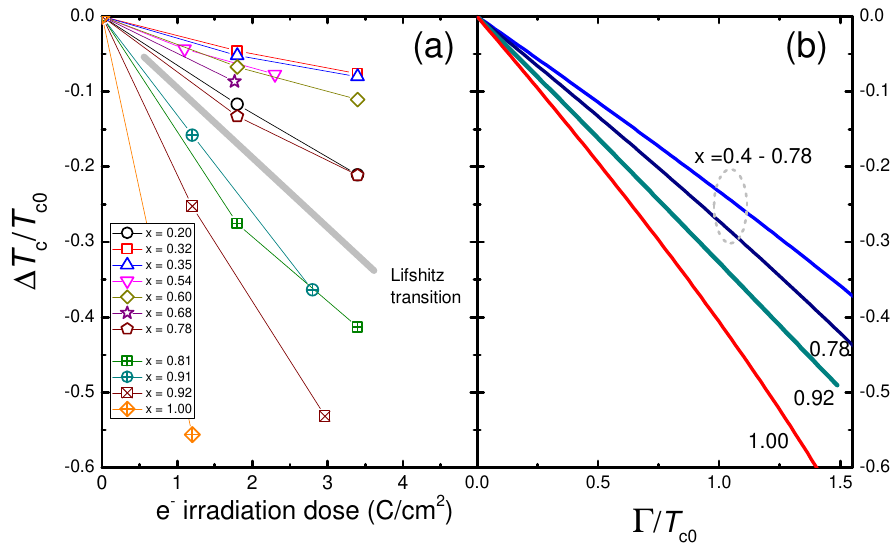}
\caption{(Color online) Variation of superconducting critical temperature upon irradiation for different compositions. (a) the normalized rate of the $T_c$ suppression versus irradiation dose. The rate increases drastically above the Lifshitz transition. (b) $t-$matrix calculations of the $T_c$ change using parameters extracted from the London penetration depth fits, \fref{fig3SI}. While we cannot expect quantitative agreement for our simplified model, the trend is clearly in line with experimental observations.}
\label{fig4SI}
\end{figure}

Here $V_{ij}$ denotes the interaction between $i^{th}$ and $j^{th}$ band. After finding the interaction parameters, impurity scattering is treated within self-consistent $t$-matrix approximation \cite{MVHV2009,MGH2011}. Before the Lifshitz transition, we consider a moderate interband scattering. The ratio between the interband and the intraband impurity potentials is $0.6$. We fix this in order to obtain the best fit. After the Lifshitz transition, interband scattering involves two concentric hole pockets. This involves a small momentum transfer, hence
after the Lifshitz transition, we take equal strengths for the interband and intraband impurity scattering potentials.
\Fref{fig3SI} shows the fits for different doping levels with the experimental data. The effect of disorder on $T_c$  within
this model for the same parameters used for penetration depth fitting is shown in \fref{fig4SI} panel (b). Note,
for fitting low temperature penetration depth a minimal two band is sufficient, but for quantitative explaining $T_c$, a full multiband approach with realistic Fermi surfaces is required.

\begin{figure}[htb]
\includegraphics[width=9cm]{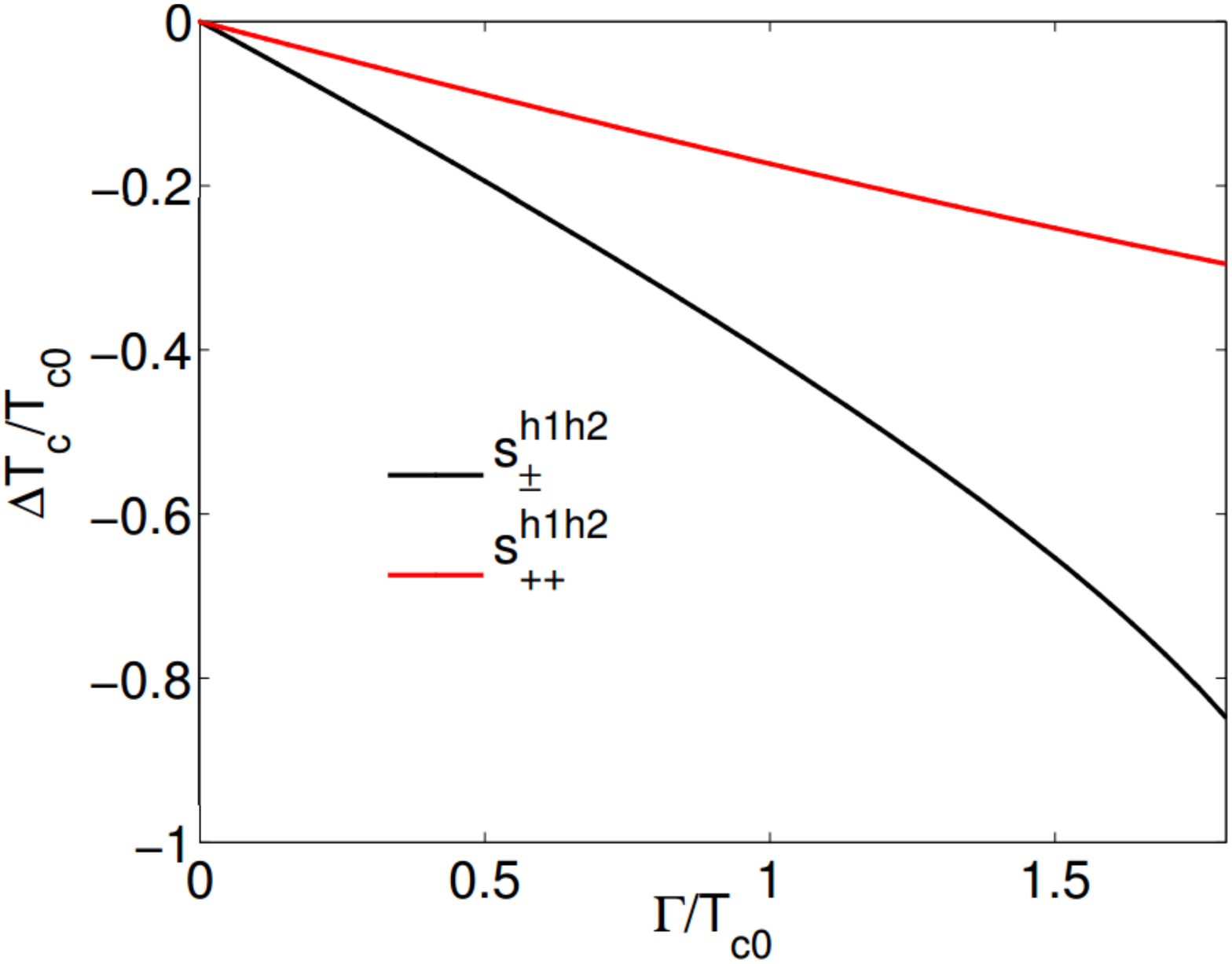}
\caption{(Color online) Comparison of $T_c$ suppression as a function of increasing disorder for various possible scenarios for heavily overdoped systems.}
\label{fig5SI}
\end{figure}

\end{document}